\documentclass[journal]{IEEEtran}
\usepackage[T1]{fontenc}

\usepackage{subfigure}
\usepackage{graphicx,url}
\usepackage{amsmath}
\usepackage{amssymb}
\usepackage{algpseudocode}
\usepackage{algorithm}
\algnewcommand\algorithmicto{\textbf{to}}
\algrenewtext{For}[3]%
{\algorithmicfor\ #1 $\gets$ #2 \algorithmicto\ #3 \algorithmicdo}

\usepackage{setspace}
\usepackage[normalem]{ulem}
\usepackage{color}
\usepackage{multirow}

\makeatletter
\newcommand\notsotiny{\@setfontsize\notsotiny\@vipt\@viipt}
\makeatother

\begin{document}
%
\title{Dynamic Multi-Modulation Allocation Scheme for Elastic Optical Networks}
%
%
%

\author{Lucas R. Costa~and~Andr\'e C. Drummond
\thanks{Lucas R. Costa and Andr\'e C. Drummond are with the Department
	of Computer Science, University of Brasilia (UnB), 70910-900, Brasilia, Brazil
	e-mail: lucasrc.rodri@gmail.com, andred@unb.br.}
}

\renewcommand{\today}{\ifcase \month \or January\or February\or March\or %
April\or May \or June\or July\or August\or September\or October\or November\or %
December\fi, \number \year} 

%



\maketitle

\begin{abstract}
In order to deal with the recent rapid increase in Internet traffic, a transmission technology is required to enable the efficient use of the optical fiber spectrum while offering flexibility in network bandwidth.
To meet these challenges, the emergence of Elastic Optical Networks (EON) has brought new conceptions in the operation of optical networks, improving flexibility and efficiency for the next generation core networks.
In EON, traffic demands are typically supported by allocating bandwidth-variable optical channels with heterogeneous modulation formats in a spectral-efficient manner.
Elastic optical path networks require the routing, modulation level, and spectrum allocation (RMLSA) to efficiently allocate optical spectrum resources to optical paths.
To address the RMLSA problem, Modulation Scheme approaches have recently been proposed to allow the use of any routing and spectrum assignment (RSA) algorithm to solve the RMLSA problem.
In this paper, we propose a new Modulation Scheme that enables the routing of traffic through dynamic multi-modulation allocation in multiple hops to achieve blocking performance improvement.
Numerical results demonstrate that the proposed adaptive modulation scheme achieves a reduction in bandwidth blocking of up to two orders of magnitude in an underloaded network scenario, and $86\%$ with higher loads, playing an important role in spectrum savings compared with the literature schemes.	
\end{abstract}


%
\IEEEpeerreviewmaketitle

\section{Introduction}

\IEEEPARstart{T}{he} volume of Internet traffic has been exponentially increasing triggered by emerging services such as high-definition video streaming, cloud computing, and inter-datacenter transmissions. 
%
Due to these characteristics, today Internet traffic is demanding different granularity levels with more increasingly flexible bit rates, provided by geographically dispersed data centers.
The consequence in the near future is that the network operators will require a new transmission technology to serve this huge and heterogeneous volume of traffic in an efficient and scalable manner~\cite{Imran}.

Driven by optical coherent detection, advances in optical transmission technologies have significantly improved the transmission rate~\cite{ofdm}.
The rigid granularity of traditional fixed-grid wavelength-division multiplexing (WDM) networks can not withstand the increased traffic and mobility of their sources~\cite{Gerstel}. 
Adapting this technology to the recent demands is one of the challenges of the Future Internet. New concepts in optical network operations are needed to improve resource efficiency and allow traffic with different granularities and flexible bit rates.

O-OFDM (Optical Orthogonal Frequency-Division Multiplexing) has been considered a promising candidate as the future high-speed optical transmission technology~\cite{Zhang}.
Faced with this, an OFDM-based Elastic Optical Network (EON) can dynamically adjust features such as optical bandwidth and modulation format, according to the requirements of each demand, improving spectral efficiency compared with fixed-grid WDM networks~\cite{Gerstel}.
By adopting finer granularity (e.g. 12.5 GHz) OFDM technology, implemented in EONs, allows a multiple-carrier modulation scheme that splits a data stream into a large number of sub-streams by establishing super-channels which provide an adaptable bit rate to ideally satisfy the band requirements, creating channels with the bandwidth required by the data flows to be transmitted~\cite{Ioannis}.
%
Compared with traditional wavelength-division multiplexing (WDM) networks, EONs require more sophisticated bandwidth allocation mechanisms, based on routing and spectrum assignment (RSA) approaches and variable bandwidth devices, such as Bandwidth-Variable transponder (BVT) and Bandwidth-Variable Wavelength Cross-Connects (BV-WXC)~\cite{Chatterjee}.

The RSA approaches are the allocation strategies for satisfying the call requests in EONs. The RSA objective is to find a path and a continuous and contiguous amount of spectrum slots~\cite{Ioannis}.
To further improve spectrum utilization efficiency, EON literature has incorporated quality-of-transmission (QoT) or transmission reach awareness into RSA solutions. This has turned the RSA approaches into routing, modulation level and spectrum assignment (RMLSA) approaches, which includes the attribution of the modulation format considering the transmission distance~\cite{Christodoulopoulos}.
In RMLSA approaches the spectral bandwidth can be saved by adopting higher modulation level if the QoT requirement is satisfied~\cite{Jinno2}.
Therefore, the RMLSA approaches becomes a fundamental problem for serving traffic requests in EON.

Several works deal with distance-adaptive allocation to address with the RMLSA problem. These studies employ the RMLSA in different network scenarios such as single-hop or multi-hop routing and static or dynamic traffic.
Recent studies on the EON explore the RMLSA problem with different Modulation Schemes~\cite{Wan2,AMMS,YLiu}. 
These schemes are proposed to allow the use of any RSA to solve the RMLSA problem by meeting traffic requests with appropriate modulation levels based on the physical distance of the routes.
%

Considering the current state of the art of EON Modulation Schemes presented in the literature, this paper proposes a new distance-adaptive Modulation Scheme to address the RMLSA problem.
Our proposal decomposes the RMLSA problem allowing the routing of traffic through multiple hops in the virtual topology (VT) maintaining network state awareness to achieve blocking performance improvement, providing less spectrum usage through the use of more efficient modulation levels.
This proposal advances in relation to our previous solution~\cite{AMMS} in several ways.
%
The new scheme dynamically evaluates the resource usage of the network to define the limit on the number of segments used in the VT to serve a traffic demand.
In addition, a new approach is proposed within the scheme to use different modulation levels to meet the same connection demand, using modulation levels appropriate to the size of each segment.
%
The general idea behind the proposed modulation scheme is to break an end-to-end route in a multi-hop path, trying to meet the most appropriate modulation level in each segment, while taking into consideration the spectrum availability.

Simulation results demonstrate that the proposed scheme achieves a gain in the bandwidth blocking rate up to three orders of magnitude in an underloaded network scenario, and $86\%$ with higher loads, playing an important role in spectrum savings compared with the literature schemes.
The EON Modulation Schemes open up a new avenue challenges for Software-defined optical network (SDON)~\cite{SDON} and outline the directions and possibilities to be explored in future research, enabling new RMLSA solutions tailored to the needs of the optical transport network providers.

The rest of the paper is organized as follows. Section~\ref{sec:concepts} introduces elastic optical network concepts, such as architecture and their elements. 
Section~\ref{sec:related_work} presents the approaches, techniques, and the state-of-the-art in EON Modulation Schemes reported in the literature. 
Section~\ref{sec:AMMS2} shows the proposed Modulation Scheme approach for addressing the RMLSA problem.
Section~\ref{sec:pe} presents numerical results of the proposed application. 
Finally, concluding remarks are given in Section~\ref{sec:conclusion}.\vspace{-0.5em}

\section{Elastic Optical Networks}\label{sec:concepts}

Elastic optical networks (EONs) enters as a promising technology for the future of high-capacity networks. Its features provide flexibility and superior scalability in spectrum allocation following the growing demands of Internet traffic. 
In this section we first introduce the basic concepts of EONs and what differs it from traditional optical networks. Then we briefly introduce the EON architecture and their devices. 
For a complete picture we present the allocation strategies for satisfying the call requests and the continuity, contiguity and transmission distance constraints of RSA/RMLSA problems.
Finally we present concepts like traffic grooming, multi-hop routing, spectrum fragmentation and introduce the energy consumption model used in this work.

\subsection{Fixed and Mini Flexible Grid}\label{sec:2-1}

EONs are characterized for dividing spectral resources into frequency slots in the form of OFDM sub-carriers, allowing multiple modulation formats and different data rates and spectrum size.
An EON is able to allocate a demand on an optical path with bandwidth appropriate to the request.
The EON optical path can be expanded or contracted as needed, according to traffic fluctuations or new connection demands. 
On traditional wavelength-division multiplexing (WDM) networks, the transmission channels use fixed-size frequency grids (50 GHz) according to International Telecommunication Union Telecommunication Standardization Sector (ITU$-$T)~\cite{Ioannis}.
These specifications accommodates 80 channels, by dividing the 4 THz C-band spectrum into 50 GHz slots. Thus, each WDM channel receives a center frequency ``$f$'' defined such that the channel occupies the spectrum between $f - 25$ GHz and $f + 25$ GHz.
So, for each traffic demand an entire grid must be allocated for its accommodation, even if it is less than the capacity of a wavelength causing inefficient use of the spectral resources.
For optimum utilization of spectral resources and to accommodate heterogeneous bandwidth demands, the ITU$-$T has extended its recommendations allowing for a mini flexible division of the optical spectrum and defining the concept of a frequency slot, in addition to a nominal central frequency and a channel/carrier spacing.
In EONs the frequency slot is defined by its nominal central frequency and its slot width. The EON channel is defined as $193.1 + n \times 0.00625$ THz, where $n$ is a positive or negative integer. The slot width is defined as $12.5$ GHz $\times$ $s$, where $s$ is a positive integer. The central frequency of the channel identifies its position in the spectrum and the slot width determines the spectrum occupied~\cite{Imran}.
\begin{figure}[!t]
	\begin{center}
		\includegraphics[width=0.5\textwidth]{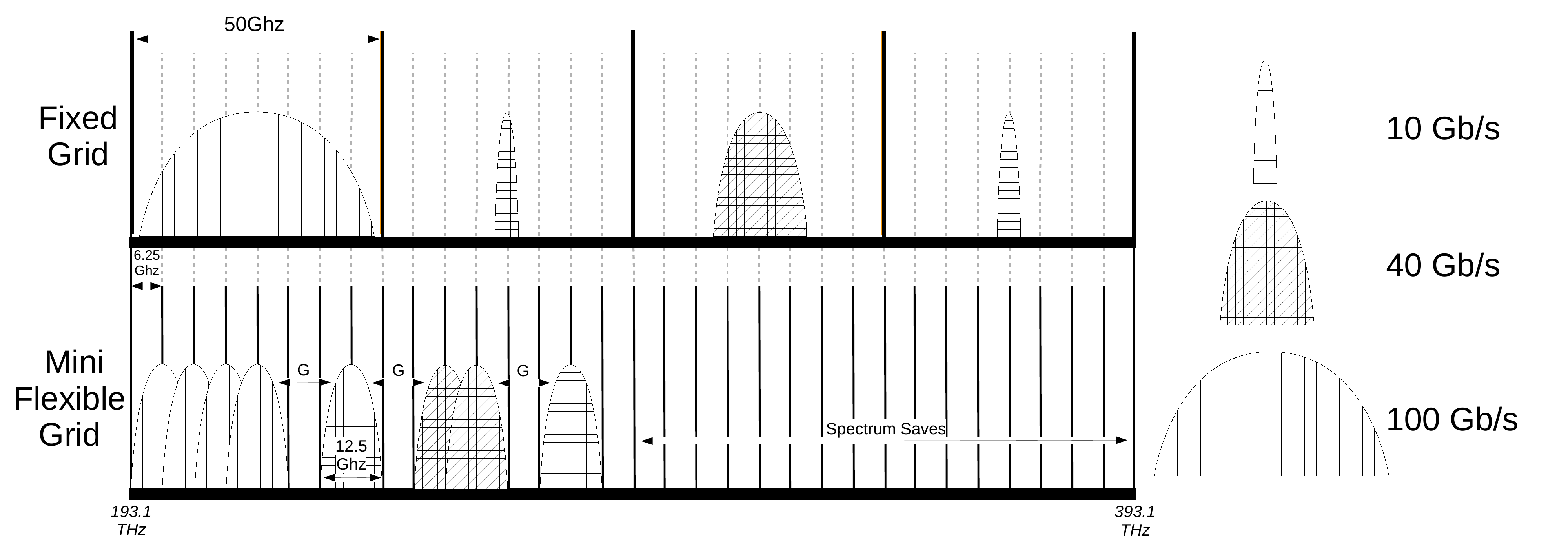}
		\caption{Fixed grid vs Mini flexible grid.}\vspace{-4ex}
		\label{fig:fixedVsflex}
	\end{center}
\end{figure}
Figure~\ref{fig:fixedVsflex} shows the differences between optical channels with fixed and mini flexible grid.
In this example, the fixed grid channel can accommodate connection requests with rates up to $100$ Gb/s.
One can observe the waste of spectral resources when the requested demands are smaller than the accommodation channel, such as $10$ Gb/s.
On the other hand, the channel with mini flexible grid offers the ability to adapt its transmission capacity, allowing the allocation of spectrum according to traffic demand.
Each channel consists of $n$ OFDM 12.5 GHz subcarriers separated by a guard band ($G$).

\subsection{EON Architecture}\label{sec:2-2}

The architecture of an EON is composed of bandwidth-variable transponders (BVTs) and bandwidth-variable Optical Cross-Connects (BV-OXC), which enable optical channels, called lightpaths, in mini flexible grids to be established.
Several slots can be joined into a channel, transporting data without any guard bands. So the BVTs create the lightpaths with flexible bandwidth, allowing the resources to be adjusted to the current demand~\cite{Zhang}. Since an elastic lightpath is allocated as required, it can transmit multiple bit rates, like $10$ Gb/s, $40$ Gb/s, $100$ Gb/s, $400$ Gb/s and $1$ Tbps.
The BV-OXCs are responsible for establishing an end-to-end path with enough bandwidth to accommodate the spectral resources defined by the BVTs.
Micro-Electro-Mechanical-Systems (MEMS) or Liquid Crystal-based (LCoS) wavelength-selective switches (WSSs) can provide bandwidth variable switching functionality to realize BV-OXCs~\cite{Christodoulopoulos}.
Thus when the BVTs increase traffic, each BV-OXC in the route must expand its switching window, allowing a variable data rate in each lightpath~\cite{Zhang}.
%
	%
%
%
%
\vspace{-2ex}
\subsection{Spectrum Representation}\label{sec:2-3}

In OFDM, data is transmitted over multiple orthogonal sub-carriers. This brings unique benefits in terms of spectral efficiency, by allowing the spectrum of adjacent sub-carriers to overlap thanks to their orthogonal waves. 
Elastic lightpaths are enables by allocating a variable number of low-rate sub-carriers to a transmission. Hence, more adjacent sub-carriers represents greater bandwidth capacity of the channel.
The modulation format used in each sub-carrier of EONs also allows flexible adjustment of bandwidth.
Since every lightpath is composed of an arbitrary number of OFDM sub-carriers, each can be individually modulated (with a different BVT) for a transmission. 
For example, single bit per symbol \textit{binary phase shift keying} (BPSK), QPSK ($2$ bits per symbol), 8QAM ($3$ bits per symbol), 16QAM ($4$ bits per symbol), etc. 

The number of sub-carriers and the modulation format are adjusted to the amount of traffic and optical reach requested, e.g., according to Table~\ref{tb:1}, a lightpath with $4$ slots in BPSK transmits $50$ Gb/s of bandwidth ($4 \times 12.5$ Gb/s). On the other hand, a lightpath with $2$ slots in QPSK have the same capacity~\cite{Zhang}.
The choice of the modulation level has to take into account the required Quality of Transmission (QoT) of the connection.
A common assumption in EON literature is that the transmission distance (or equivalently the number of traversed spans for equal spaced spans) of the lightpath is claimed to be the most relevant factor in QoT~\cite{Christodoulopoulos, Jinno2}.
Therefore, given the length of the lightpath, the modulation level that provides the best spectrum efficiency without hindering the QoT can be found. 
This allows shortest lightpaths to use higher modulation levels as can be seen in Table~\ref{tb:1}.
Note that the maximum transmission distance of each modulation format differs widely in the literature and possible signal regenerators further extend the tolerable reach.
In addition, modulation's reach follows the half distance law that indicates the maximum transmission distance decreases by half as the modulation level increases by one~\cite{Christodoulopoulos, Wan2}.
Also note that other parameters related to BVTs (transmitter/receiver) characteristics, interference and non-linear physical layer impairments, can also affect the QoT and thus the modulation level choice. 
However, this scenario is outside the scope of this paper.
%
\begin{table}[!h]
	\footnotesize
	\centering
	\caption{Spectrum size correlation, bandwidth, maximum transmission distance and energy consumption of ODFM sub-carriers.}
	\label{tb:1}
	\begin{tabular}{cccc}
		\hline
		\multicolumn{1}{c}{\textbf{\begin{tabular}[c]{@{}c@{}}Modulation \\ Format\end{tabular}}} & \multicolumn{1}{c}{\textbf{\begin{tabular}[c]{@{}c@{}}Subcarrier \\ Capacity (Gb/s)\end{tabular}}} & \multicolumn{1}{c}{\textbf{\begin{tabular}[c]{@{}c@{}}Transmission \\ Distance (km)\end{tabular}}} 
		& \multicolumn{1}{c}{\textbf{\begin{tabular}[c]{@{}c@{}}Power \\ Consumption (W)\end{tabular}}}\\ \hline
		BPSK                                                                                          & 12.5                                                                                                      &  8000   &112.374                                                                                       \\ 
		QPSK                                                                                          & 25.0                                                                                                        &  4000 &133.416                                                                                         \\ 
		8QAM                                                                                          & 37.5                                                                                                      &  2000      &154.457                                                                                    \\ 
		16QAM                                                                                         & 50.0                                                                                                        &  1000     &175.498                                                                                     \\ 
		32QAM                                                                                         & 62.5                                                                                                      &  500           &196.539                                                                                \\ 
		64QAM                                                                                         & 75                                                                                                        &  250          &217.581                                                                                 \\ 
		\hline
	\end{tabular}\vspace{-4ex}
\end{table}
%
\subsection{EON Traffic Engineering}\label{sec:2-4}
\begin{figure}[!t]
	\begin{center}
		\includegraphics[width=0.45\textwidth]{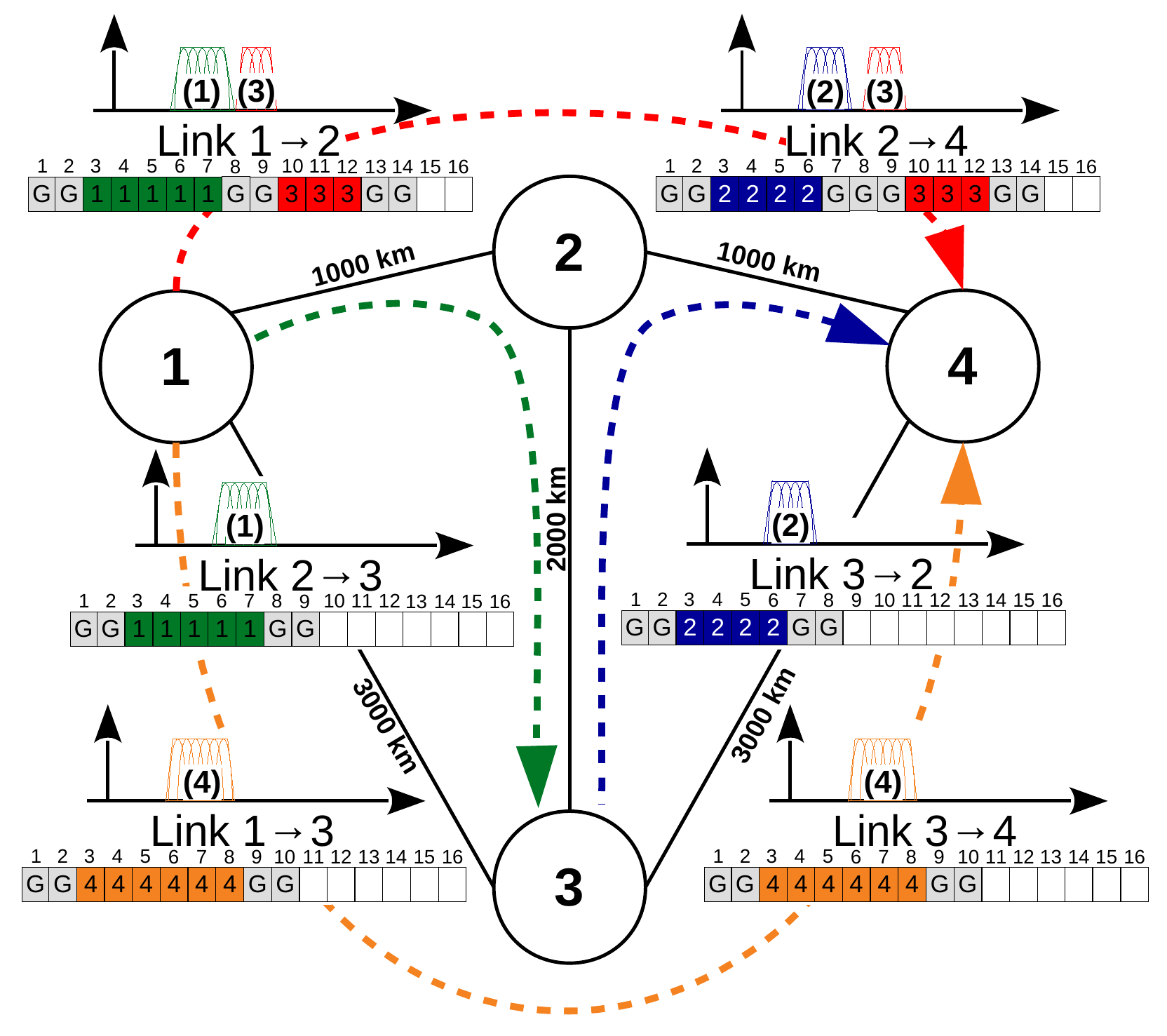}\vspace{-3ex}
		\caption{Traffic Engineering in a four-node EON.}\vspace{-5ex}
		\label{fig:rsa}
	\end{center}
\end{figure}
The lightpaths are established by means of traffic engineering (TE) algorithms executed by the network control plane. 
The control plane performs the control functions such as configuration, establishment or termination of the connections.
In addition, the control plane needs to know the network state and the limitations of network resources while controlling traffic flows and aims to maximize network performance~\cite{Ioannis}.

The routing and spectrum assignment (RSA) problem requires sub-carriers in the same optical path must be routed contiguously using the same spectrum band throughout the route, and adjacent lighpaths must be spaced by a guard band to avoid interference effects and attend OFDM restrictions.
The RSA limitations are illustrated in Figure~\ref{fig:rsa}, which considers an EON with four nodes and $4$ established lightpaths in the network. 
In this example, the first lightpath, denoted by $(1)$, is using five slots and being routed through nodes $1;2;3$, using the same spectrum band throughout the route (3-7 slots; 1$\rightarrow$2 and 2$\rightarrow$3 links). The same happens with lightpath $(2)$, but their using four slots and consequently transmitting smaller bandwidth than the previous one.
Adjacent optical paths are separated by a minimum two-slots guard band, denoted by~G.
%

If there is additional flexibility in the modulation format selection, then routing, modulation level, and spectrum allocation (RMLSA) techniques are in order~\cite{Christodoulopoulos}. 
This introduces an extra degree of flexibility in TE operations in the EONs and at the same time a greater complexity in the spectrum allocation problem.
In RMLSA, each demand is mapped to a modulation level based on the requested data rate and the distance of the path over which it is routed.
The main difference between RSA and RMLSA is that the spectrum allocation model is used to determine the number of sub-carriers as a function of data rate and path length.
In this sense, still taking as an example, Figure~\ref{fig:rsa} and the considerations of Table~\ref{tb:1}, considering lightpath, denoted by $(3)$, using three slots and routed through nodes $1;2;4$ ($2000$ km path length), the QPSK modulation level can be used to transmit a data rate of $75$ Gb/s.
On the other hand, taking route $1;3;4$ ($6000$ km path length), such as lightpath denoted by $(4)$, are necessary six slots to transmit the same data rate as lightpath $(3)$, because it uses the BPSK modulation to meet the transmission distance constraint.
Thus, although the lightpath $4$ has more slots than the lightpath $3$, its transmission capacity is the same.\vspace{-2ex}

\subsection{Another Traffic Engineering Considerations}\label{sec:2-5}

The EON performance depends not only on its physical resources, such as transponders, physical links, usable spectral width, optical switches, etc., but also on how the network is controlled.
The objective of a TE solution is to achieve the best performance within the limits set by the physical constraints and can be cast in numerous forms.

\subsubsection{Traffic Grooming}
One of the issues of these algorithms is satisfying traffic demands with low bit rates. 
Though BVTs can dynamically adjust the offered bandwidth in an EON, if the demand needs less bandwidth than the available in the OFDM subcarrier, the spectrum will be underused. Additionally, several such demands might produce a significant amount of guard bands, leading to squandering of the spectrum.
To address these problems, a traffic grooming approach in EON was proposed in~\cite{Zhang2}, in which multiple low-speed traffic requests are groomed into elastic optical paths using electrical layer multiplexing. Electrical traffic grooming also improves the resource utilization by aggregating multiple channels (packet or circuit flows) onto a single optical channel.
%

\subsubsection{Multi-hop routing}
A TE algorithm can be classified as single-hop or multi-hop. Single-hop when a connection request is served by only one lightpath. Multi-hop when a connection request goes through several lightpaths, keeping in mind each lightpath can span several links and regenerators.
At the beginning/end of each lightpath, an OEO (Optical-Electrical-Optical) conversion will be performed by a BVT, which will result in a hop in the virtual topology.
The lightpaths considered, may be new or already established. New when lightpaths were created to serve a new connection and existing when lightpaths are used by traffic grooming.
The multi-hop routing relax the spectrum continuity constraints because the lightpaths of the solution do not need to have the same spectrum segment among them.
In addition, this kind of routing minimizes the transmission distance constraints, since the transmission is divided into several lightpaths and the signal is completely regenerated~\cite{MBM}.

\subsubsection{Fragmentation}
Under a dynamic traffic scenario, the process of establishing and releasing connection requests, inevitably creates small non-contiguous spectrum fragments, leading to the so-called spectrum fragmentation problem~\cite{Zhang}.
This problem causes the inefficient use of resources, since most of the future requests are not met due to a lack of contiguous spectrum available, thus causing a degradation of network performance.
The TE algorithm adopted by the dynamic scenario must take these issues into account in order to perform satisfactorily.
There are several metrics to measure the fragmentation ratio.
One of the best known and adopted in the literature~\cite{Rosa} is the external fragmentation ratio, commonly used in computer architecture, and denoted by Eq.~\ref{eq:fextRosa}:
\begin{equation}~\label{eq:fextRosa} 
    \small F_{ext} = 1 - \dfrac{LargestFreeBlock}{TotalFreeSlots}
\end{equation}
\noindent
where $LargestFreeBlock$ represents the number of slots of the largest contiguous free space, and $TotalFreeSlots$ is the total number of slots available. If $F_{ext}$ is close to one, it means that the space available on the channel is all broken into small pieces.
To measure the network fragmentation degree the average of $F_{ext}$ is made for all links.
In~\cite{WrightFrag} the authors propose a new metric to measure the fragmentation ratio using the Shannon entropy, in which each link is evaluated according to Eq.~\ref{eq:fragWright}:
\begin{equation}~\label{eq:fragWright} 
    \small F_{ent} = - \sum_{i=1}^{N}\dfrac{D_i}{D}\ln\dfrac{D_i}{D}
\end{equation}
\noindent
where $N$ indicates the number of free sub-blocks, $D_{i}$ represents the size of sub-block $i$ and $D$ represents the number of slots in the spectrum.
To measure the network fragmentation degree the sum of $F_{ent}$ is calculated for all links.
It is important to highlight that this evaluation prioritizes the free blocks, regardless of the distribution of them in the links.
Therefore, the greater the value, the greater the fragmentation.
In this work, we use Eq.~\ref{eq:fextRosa} to display the performance evaluations and Eq.~\ref{eq:fragWright} as one of the metrics used by the proposal.

\subsection{Energy Consumption}\label{sec:2-6}

Energy consumption and efficiency are becoming major concerns for network planning/operation.
It is important to build network infrastructures that enable energy savings that take into account economic relations, as well as ecological consequences~\cite{Zhang}.
TE solutions should also take into account the energy expenditure in their decision-making aiming to provide energy efficiency in the network.

Due to the commercial unavailability of OFDM (BVT / BV-OXC) variable bandwidth components some hypotheses have been made in the literature to measure the energy expenditure of EON components~\cite{Rongrong, VizcainoEnergy}.
The power consumption of a single BVT can be interpolated as function of its transmission rate (TR) as shown in Eq.~\ref{eqn:bvt}.
The energy consumption of the BVTs for a subcarrier in the different modulation formats is shown in Table~\ref{tb:1}.
%
Furthermore, an additional $68.3\%$ is considered for possible overloads plus $91.333$ $W$ for idle consumption in each BVT.
\begin{equation}\label{eqn:bvt}
	\small PC_{BVT}(W)= 1.683\times TR(Gb/s)+91.333
\end{equation} 
The energy consumption of the BV-OXCs comes from the setup and operation procedures. The setup procedure depends on the node degree $N$ (the number of fibers connected to the node) and the add/drop degree $\varepsilon$ (the number of channels which can be added or dropped locally), as shown in Eq.~\ref{eqn:oxc}.
\begin{equation}\label{eqn:oxc}
	\small EC_{OXC}^{setup}(J)= N \times 85 + \varepsilon \times 100
\end{equation} 
The power consumption of the operating procedure is $150$ $W$ for nominal overhead contribution~\cite{Rongrong, VizcainoEnergy}, thus $PC_{OXC}^{op}(W)= 150$.
In relation to optical inline amplification (OLA) Erbium Doped Fiber Amplifiers (EDFAs) are considered, and their power consumption is 100 W for each 80Km span. Thus $PC_{OLA}(W)= 100$.
Therefore, the total energy consumption to transmit a lightpath depends on these elements and their holding time ($H_{LP}$), as shown in Eq.~\ref{eqn:lp}.
\begin{equation*}
	\small
	EC_{LP}(J) = \sum_{k=1}^{C} {EC_{OXC}^{setup}}_{k} 
\end{equation*} 
\begin{equation}\label{eqn:lp}
	\small
	+ \left( PC_{BVT} + \sum_{i=1}^{C} {PC_{OXC}^{op}}_{i} + \sum_{j=1}^{A} {PC_{OLA}}_{j} \right) \times H_{LP}
\end{equation} 
\noindent
where, $C$ indicates the number of BV-OXCs used in the lightpath and $A$ is the number of OLAs used.

The data transmitted is given by the product of the transmission rate of the flow request (TRFlow) and the flow holding time ($H_{Flow}$), as shown in Eq.~\ref{eqn:flow}.
\begin{equation}\label{eqn:flow}
	\small DT_{Flow}(bits)= TRFlow(bit/s) \times H_{Flow}(s)
\end{equation} 
Finally, the total energy consumption of the network is calculated by adding the energy consumption of all lightpaths allocated in the network (Eq.~\ref{eqn:lp}).
The total data transmitted is obtained by summing the data successfully transmitted in the different flows requests (Eq.~\ref{eqn:flow}).
Then the energy efficiency of the network (bits/Joule) is defined as the ratio between the total data transmitted, and the total energy consumed in the network, as shown in Eq.~\ref{eqn:ee}.
\begin{equation}\label{eqn:ee}
	\small EnEff (bits/Joule) = \frac{TotalDT(bits)}{TotalEC(J)}
\end{equation} 
\section{Related Work}\label{sec:related_work}

Recent studies on EON employs several spectrum allocation approaches that use different network scenarios like single-hop or multi-hop routing and fixed or adaptive modulation.

Some heuristics for solving the RSA problem in a dynamic scenario were proposed in~\cite{Wan2}. The first is a two-step approach which initially applies Yen's algorithm~\cite{yenksp}  for computing single-source \textit{K-shortest paths} (KSP) that are then processed in an attempt to allocate the demand to one of them.
Then the authors propose two more heuristics (MSP and SPV) based on one-step RSA approaches (concurrently computed routing and spectrum allocation).
The MSP (\textit{Modified Dijkstra Shortest Path}) attempts to find the shortest spectrum assignment path.
The SPV (\textit{Spectrum-Constraint Path Vector Searching}) algorithm applies a breadth-first search to create a tree that represents the candidate paths and finds, within the options that provide the demanded spectrum, the shortest one. 

Subsequently, the authors in~\cite{Wan2} investigated the effects of these RSA algorithms when using adaptive modulation (RMLSA).
To this end, the authors proposed the \textit{m Adaptive RSA algorithms}, called \textit{\textbf{mAdap}}, which iterate through possible modulations, in decreasing order, applying the RSA algorithm until a solution is found. Experimental results for this work show significant reduction in blocking and in spectral usage.

In~\cite{Zilong} the authors propose the first distance-adaptive modulation algorithm with traffic grooming techniques. It optically groups traffic demands with the same source that share common links and commutes them with no guard bands, while considering the modulation in use in the optical tunnel.
In our previous work~\cite{MBM}, a RMLSA approach was introduced that aims to apply optical grooming as much as possible through multi-hop routing, using the most spectrally efficient modulation level, called MBM (\textit{Maximize the use of Best Modulation format}). The results were compared with several RSA approaches considering the \textit{mAdap} scheme.

Considering the advantages of multi-hop routing with the use of adaptive modulation, we present in~\cite{AMMS} a new adaptive modulation scheme, called \textbf{AMMS} (\textit{Adaptive Modulation Multihop Schema}), that establishes solutions of traffic engineering through multiple hops in the virtual topology.
The AMMS aims to assign appropriate modulation levels associated with a suitable number of hops on accepting a connection request.
Besides to transforming RSA algorithms into RMLSA, the advantage of the AMMS is to transform single-hop TE solutions into multi-hop. 
The number of hops are empirically defined by means of a control mechanism and a constant.
Simulation results showed that the AMMS may provide a reduction of up to $82\%$ in the bandwidth blocking ratio using $7\%$ less spectral resources in the network, compared with \textit{mAdap}.

In~\cite{YLiu} the authors propose the first distance-adaptive modulation scheme focusing on energy efficiency, called \textbf{EEMS} (\textit{Energy-Efficient Modulation Scheme}).
The EEMS aims to find the lowest modulation level, without increasing the spectrum resources cost.
For this, the scheme proposes a mechanism that uses more robust modulation levels without increasing the spectrum (number of slots) in order to save energy, considering that robust modulation levels are more energy efficient.
The results demonstrate that EEMS scheme does not increase the spectrum consumption while saving power consumption compared to mAdap scheme.

Based on the adaptive modulation schemes presented, it can be observed that the use of these schemes brings benefits to the TE solutions. 
Although the mAdap scheme provides the most suited modulation level to TE solution, later studies (eg~\cite{MBM, AMMS}) have shown the advantages that multi-hop brings using more efficient modulations and saving spectrum, such as AMMS.
%
%
\vspace{-1ex}
\section{Dynamic Multi-Modulation Allocation Scheme}\label{sec:AMMS2}

\begin{figure}[ht]
	\begin{center}
		\includegraphics[width=0.35\textwidth]{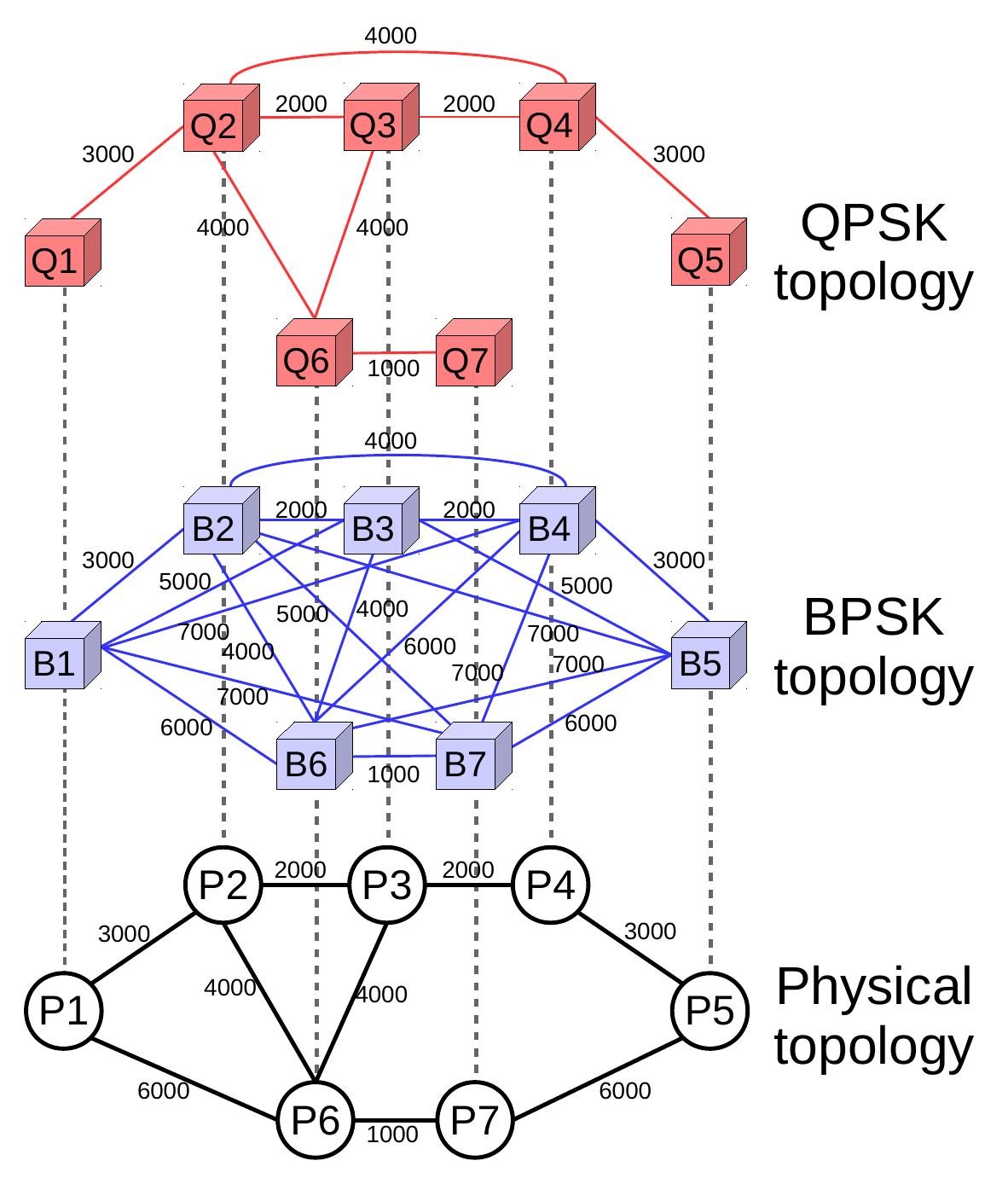}
		\caption{Virtual modulation topologies for a 7-node topology. Edge weights are set in kilometers.}\vspace{-4ex}
		\label{fig:mt}
	\end{center}
\end{figure}

Considering the current state of the art of EON Modulation Schemes presented in the literature, we propose the DMMAS (\textit{Dynamic Multi-Modulation Allocation Scheme}), a scheme that employs multiple modulation formats on the allocation of multi-hop paths without running down network resources.
%
%
%
From the physical topology the DMMAS builds (off-line) modulation topologies whose edges represent the reachability zone of each node.
Each edge in specific modulation topology is constructed using the shortest path between nodes and should also satisfy the modulation level reach to provides an acceptable QoT in network.
Figure~\ref{fig:mt} shows virtual modulation topologies for a 7-node sample physical topology.
In this example the BPSK and QPSK modulation formats were considered allowing maximum reaches of 8000 and 4000 km, respectively, according to Table~\ref{tb:1}.

It can be observed that in each modulation topology the edges are constructed from the minimal reach possible between nodes.
The edge weight is the shortest distance between nodes (the sum of edges weights which composes the shortest path).
Therefore this represents that each pair of nodes connected by an edge can establish an optical path end-to-end at certain modulation level.
For example, consider a connection request from P1 to the P4, you can accept the connection to only one optical path ($B1\rightarrow B4$) in BPSK modulation, however it requires at least two optical paths ($Q1\rightarrow Q2\rightarrow Q4$) to meet the connection with the QPSK modulation.
The general idea is to break the connection request in sub-paths, by trying to meet the most appropriate modulation level given the request.
To this end, the scheme includes a data structure, computed off-line, composed by $k$-shortest paths for each node pair in each modulation topology.
The DMMAS includes a routine $\omega(s,d,k,M)$ which gets the $k$-path considering the source and destination nodes of the connection request ($s$ and $d$), for the specific $M$ modulation topology.

\begin{figure}[ht]
	\begin{center}
		\includegraphics[width=0.5\textwidth]{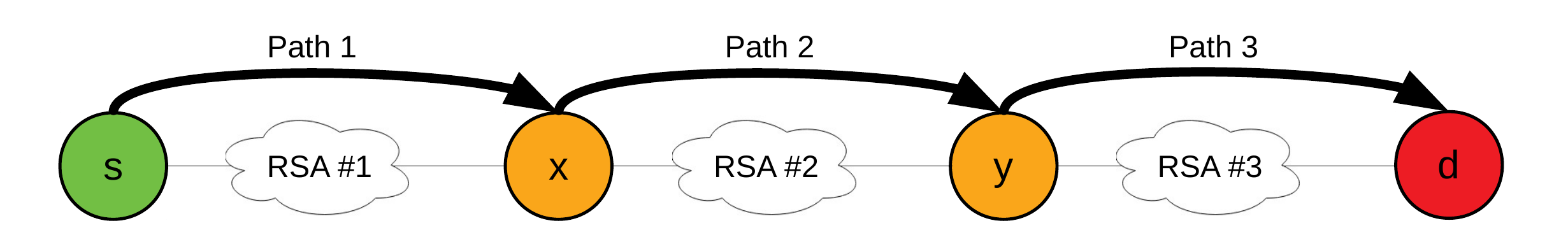}
		\caption{Relationship between articulation nodes and the number of subpaths of a set $P$ for a request $R\{s,d,b\}$.}\vspace{-4ex}
		\label{fig:mt2}
	\end{center}
\end{figure}

The $\omega(s,d,k,M)$ routine returns a set of sub-paths $P = {p_{1} m_{1}, \dots ,p_{n} m_{n}}$ \textit{st} $p_{i} = \{s',d',b\}$, where each sub-path $p_{i}$ is associated with a modulation level $m_{i}$, and each $p_{i}$ compose a part of the path between the origin and destination nodes ($s$ and $d$) of the request.
Therefore, $\omega$ defines the articulation nodes of the request ($R\{s, d, b\}$) RMLSA solution.
Each crossing in each articulation node represents an OEO conversion or, in other words, one hop in the virtual topology.
Consequently, $\omega$ returns a set $P$, where its size ($|P|$) represents the number of RSA executions required to serve the demand.
Figure~\ref{fig:mt2} illustrates the relationship between articulation nodes and the number of subpaths of a set $P$ for a request $R\{s,d,b\}$. 
Assuming that $\omega(s,d,k,m)$ returns $P = {p_{1} m_{1}, p_{2} m_{2}, p_{3} m_{3}}$, there will be 2 articulation nodes ($x$ and $y$), 3 paths ($p[s,x],p[x,y],p[y,d]$), consequently three RSA executions, and at least 4 nodes in the RMLSA solution for $R$, since each $p_{i}$ can pass through several nodes in the physical topology.
\subsection{Multi-Hop Constraint}\label{sec:mhc}

In order to choose the appropriate modulation level, a specific number of articulation nodes must be calculated, since many articulation nodes will bring up many virtual hops which may impair the quality of the solution.
In~\cite{AMMS}, a constant, called $MHC$ (\textit{Multi-Hop Constraint}), was defined based on empirical data.
For DMMAS the constant was modified to be a function of the current network utilization state.
The $MHC$ factor is a control mechanism to set the appropriate number of virtual hops in DMMAS schema. Moreover, $MHC$ also provides the choice of the modulation level for the RMLSA solution.
To set the $MHC$ factor (Eq.~\ref{eqn:mhc}), we take into account the network topology, network modulation levels and the current network state fragmentation index\footnote{$F_{ent} = \{x \in {\rm I\!R} : 0 < x \leq 1 \}$}.
\begin{equation}\label{eqn:mhc}
	\small MHC = \left \lceil{ \dfrac{\textrm{\textit{dia}} \times F_{ent} \phantom{}^{\textrm{\footnote[1]{}}}}{Reach(maxM)} }\right \rceil
\end{equation}
where ``$dia$'' is the network diameter and ``$Reach(maxM)$'' represents the network reach of the most spectrally efficient modulation available. 
Finally, we calculate the network fragmentation index based on Eq.~\ref{eq:fragWright}.
Therefore, $MHC$ defines an upper bound on the number of articulation nodes.
More specifically, $MHC$ represents the maximum number of virtual hops the DMMAS should calculate in order to accept a connection request.
If the $\omega(s,d,k,M)$ routine returns a path of which the number of nodes is larger than $MHC$, then the scheme ignores it in order to assure paths with fewer hops in the virtual topology.
Thus, the fragmentation index is used as a means to evaluate the state of the network and propose solutions with more or less hops in the virtual topology, considering the scarcity of spectrum.\vspace{-2ex}

\subsection{Spectral Efficiency Module}\label{sec:bems}

\begin{figure}[!t]
	\begin{center}
		\includegraphics[width=0.4\textwidth]{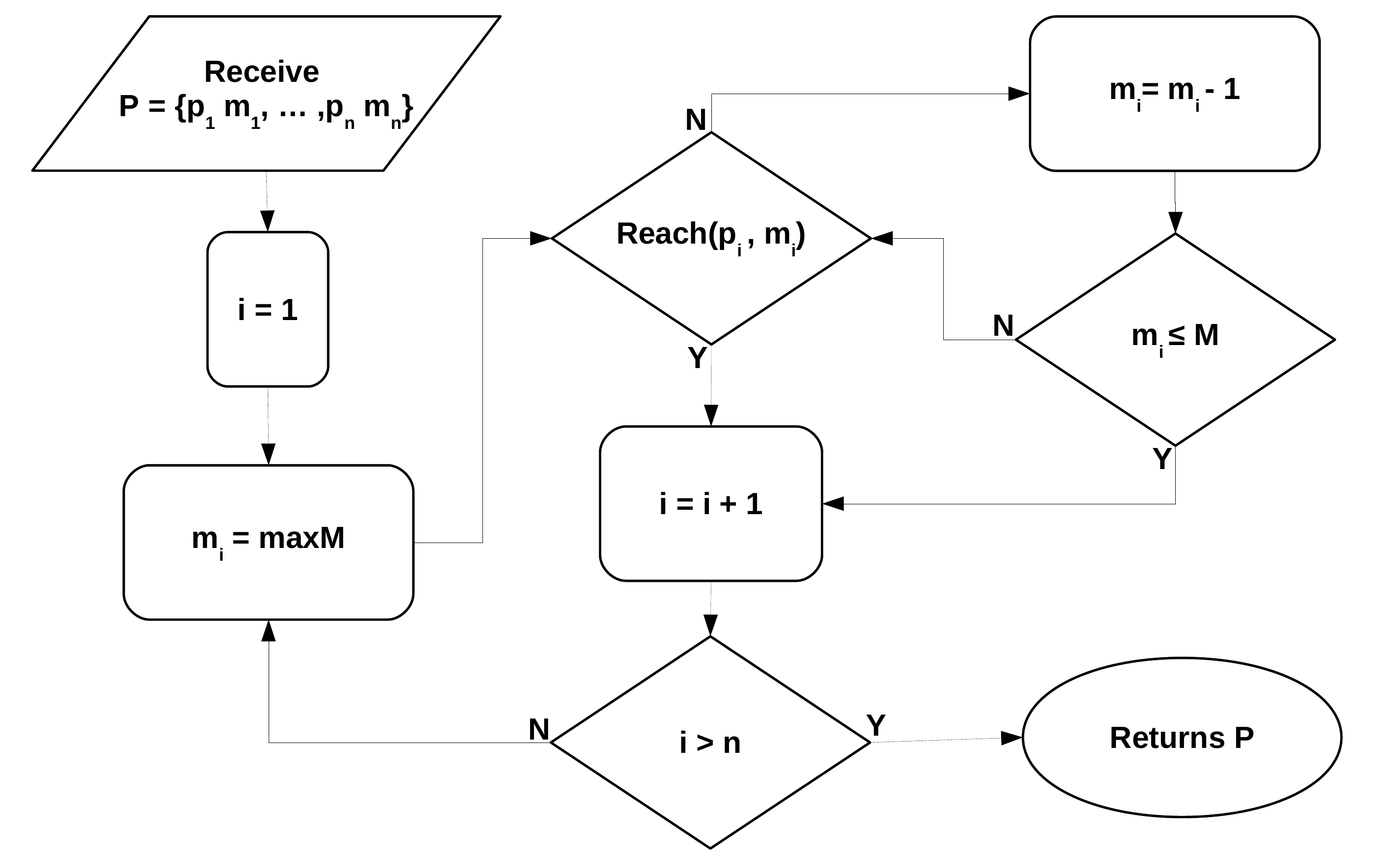}
		\caption{Spectral Efficiency Module (SpecEff-Module).}\vspace{-4ex}
		\label{fig:bem}
	\end{center}
\end{figure}

In~\cite{AMMS} all segments of the path use the same modulation level, for DMMAS multiple modulation levels can be used. In order to do that, DMMAS implements SpecEff-Module.
This module aims to find the highest level of modulation (bits per symbol) for each $p_{i} \in P$.
The SpecEff-Module module can be seen in Figure~\ref{fig:bem}.
The algorithm receives a set $P$ from the $\omega$ routine and assigns $i = 1$.
Next, $m_{i}$ receives the most spectrally efficient modulation available ($maxM$), then verifies if path $p_{i}$ meets the transmission distance constraint of modulation $m_{i}$ ($Reach(p_{i},m_{i})$), according to Table~\ref{tb:1}. 
If the transmission distance constraint is not satisfied, then decrement $m_{i}$ and check if $m_{i}$ is less than or equal to its initial modulation $M$. If not, the loop is executed again.
If the transmission distance constraint is satisfied or $m_{i} = M$, then $i$ is incremented and the same procedure is performed for the other sub-paths in $P$ until its return.
In summary, the SpecEff-Module aims to assign the best modulation levels for each $p_{i} \in P$.\vspace{-2ex}

\subsection{Modulation Scheme}\label{sec:ammsfa}

\begin{figure}[!t]
	\begin{center}
		\includegraphics[width=0.45\textwidth]{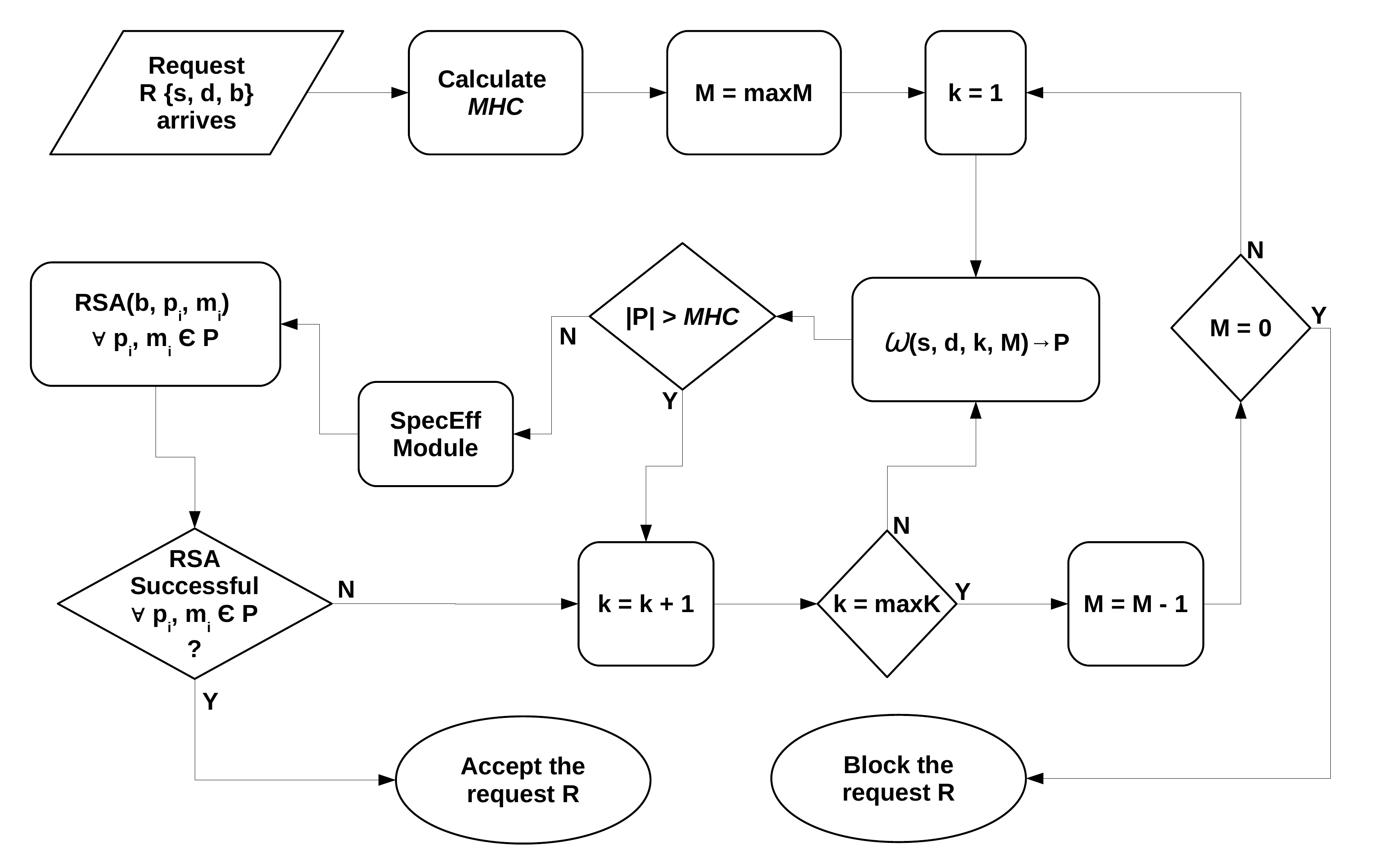}
		\caption{General outline of the proposed solution (DMMAS).}\vspace{-4ex}
		\label{fig:ammsfa}
	\end{center}
\end{figure}

The DMMAS scheme is shown in Figure~\ref{fig:ammsfa}.
When a connection request $R\{s, d, b\}$ arrives, 
the first step is to calculate MHC, to set the maximum number of hops (lightpaths) that can be used to serve $R$.
Then the schema choose the most spectrally efficient modulation available ($M = maxM$) then set $k = 1$ and runs the $\omega(s,d,k,M)$ routine, based on $k$-shortest path in $M$ modulation topology, to define the paths of the set $P$.
Subsequently, the algorithm verifies if the number of sub-paths in $P$ ($|P|$) is greater than $MHC$.
If not, the set $P$ is sent to the SpecEff-Module which assigns the most spectrally efficient modulation levels greater than $M$ for each $p_{i} \in P$.
Next, SpecEff-Module returns $P$ for the RSA module so that it can solve the RSA problem for every $p_{i} \in P$ with the modulation level $m_{i} \in P$.
If the RSA module can assign spectrum for all sub-paths in $P$, then the request $R$ is accepted in the network. Otherwise or if $|P| > MHC$ (previous step), then increments the $k$ value.
Next step checks if $k = maxK$, if not, runs the $\omega(s,d,k,M)$ routine again with new $k$ value, otherwise decrements $M$, sets $k = 1$ and runs the previous steps again. This loop iterates until $M=0$, then block the request.
\vspace{-2ex}

\subsection{Complexity Analysis}\label{sec:ca}

The time complexity of the off-line computation phase in DMMAS schema is analyzed as follows.
For constructing the modulations topologies, $|V|^2$ executions of Dijkstra's algorithm (each node pair), are required. Thus the time complexity is $O(|V|^2*(|E|+|V|log|V|))$, where $E$ is set of bidirectional links and $V$ is set of nodes in the network. For the $k$-shortest path in each modulation topology ($m$) for each node pair, Yen's algorithm~\cite{yenksp} is employed, thus the final off-line time complexity is $O(m*|V|^2*(k*|V|(E+|V|logV)))$.

The time complexity of the on-line computation phase in DMMAS schema is based on the amount of executions of the RSA algorithm considered. 
The $\omega(s,d,k,M)$ routine can be performed up to $maxK * maxM$ times and each execution provides up to $|V|$ RSA executions.
For MHC calculation, the time complexity of Eq.~\ref{eq:fragWright} is considered for each link in the network, which is given by $O(|E|*S)$, where $S$ is the number of slots of a link.
The time complexity of SpecEff-Module is $O(maxM*|V|)$.
Thus, the time complexity of the DMMAS scheme is $O(|V|*maxK*maxM*|E|*S)$ multiplied by the complexity of the RSA algorithm.
It is important to note that $maxK$ and $maxM$ are small numbers in practice.\vspace{-1ex}
\section{Performance Evaluation}\label{sec:pe}

Numerical simulations have been performed to evaluate the performance of the proposed DMMAS scheme comparing to \textit{mAdap}, AMMS and EEMS schemes considering the metrics described in the following subsections.
Moreover, in order to evince relevance of the dynamic MHC, we also consider in our analysis the DMMAS without MHC (DMMASwoMHC). Without the MHC, the scheme will have no limitation in the number of virtual hops employed.
%
For all modulation schemes, the RSA algorithm used was the classic KSP-FF RSA algorithm based on the two-step approach that partitions the RSA problem into two sub-problems, routing and spectrum assignment, and solves them iteratively.
	%
	%
	For KSP algorithm, $k=3$, the chosen policy for spectrum allocation is First Fit (FF) and performs electrical traffic grooming using the least used lightpath policy.
	%

Simulations were performed using the Optical Network Simulator\footnote{Code can be found at: \url{https://gitlab.com/get-unb/ons}} (ONS)~\cite{ONS}.
The independent replication method was employed to generate confidence intervals with $95\%$ confidence level.
Each simulation run involved $10^{5}$ requests with six types of connection requests: $25$ Gb/s, $50$ Gb/s, $100$ Gb/s, $200$ Gb/s, $300$ Gb/s, and $400$ Gb/s, with their proportion being 6:5:4:3:2:1, respectively. 
Connection requests follow a Poisson process with the mean holding time of 600 seconds, according to a negative exponential distribution and uniformly-distributed among all nodes-pairs.
The granularity of frequency slot is $12,5$ GHz with a total of $320$ slots in each fiber. The guard band between two adjacent lightpaths is assumed to be 2 slots.
Each node is a BV optical cross-connect (BV-OXC) equipped with sufficient BV-transponders, each with a maximum capacity of $32$ slots.
The modulation formats considered were BPSK, QPSK, 8QAM, 16QAM, 32QAM and 64QAM with $1$, $2$, $3$, $4$, $5$, and $6$ bits per symbol, respectively. Each modulation's reach follows the half distance law, and the maximal distances follows Table~\ref{tb:1}.

\begin{figure}[ht]
	\begin{center}
		\includegraphics[width=0.3\textwidth]{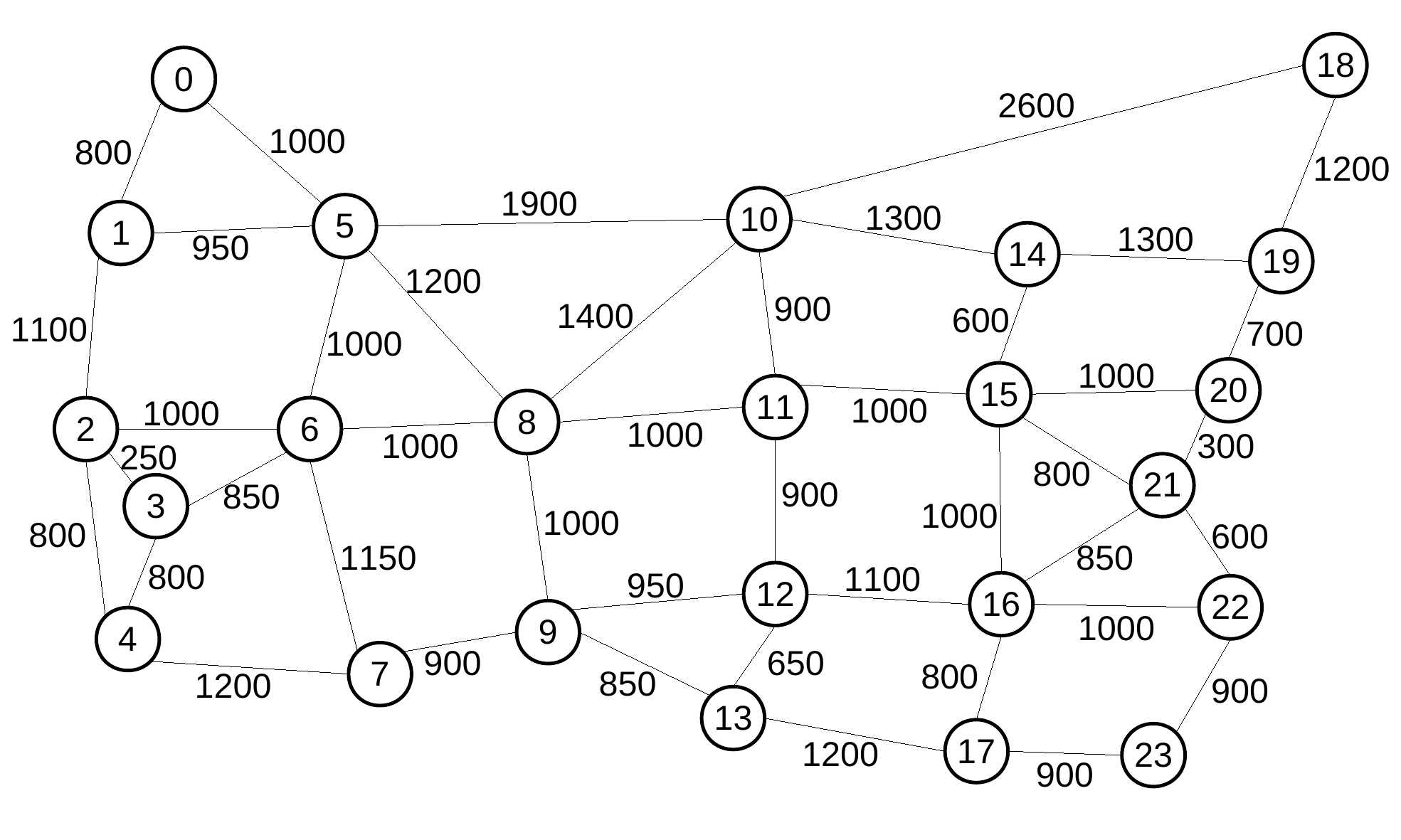}
		\caption{USA topology.}\vspace{-4ex}
		\label{fig:usanet}
	\end{center}
\end{figure}

\begin{figure}[ht]
	\begin{center}
		\includegraphics[width=0.3\textwidth]{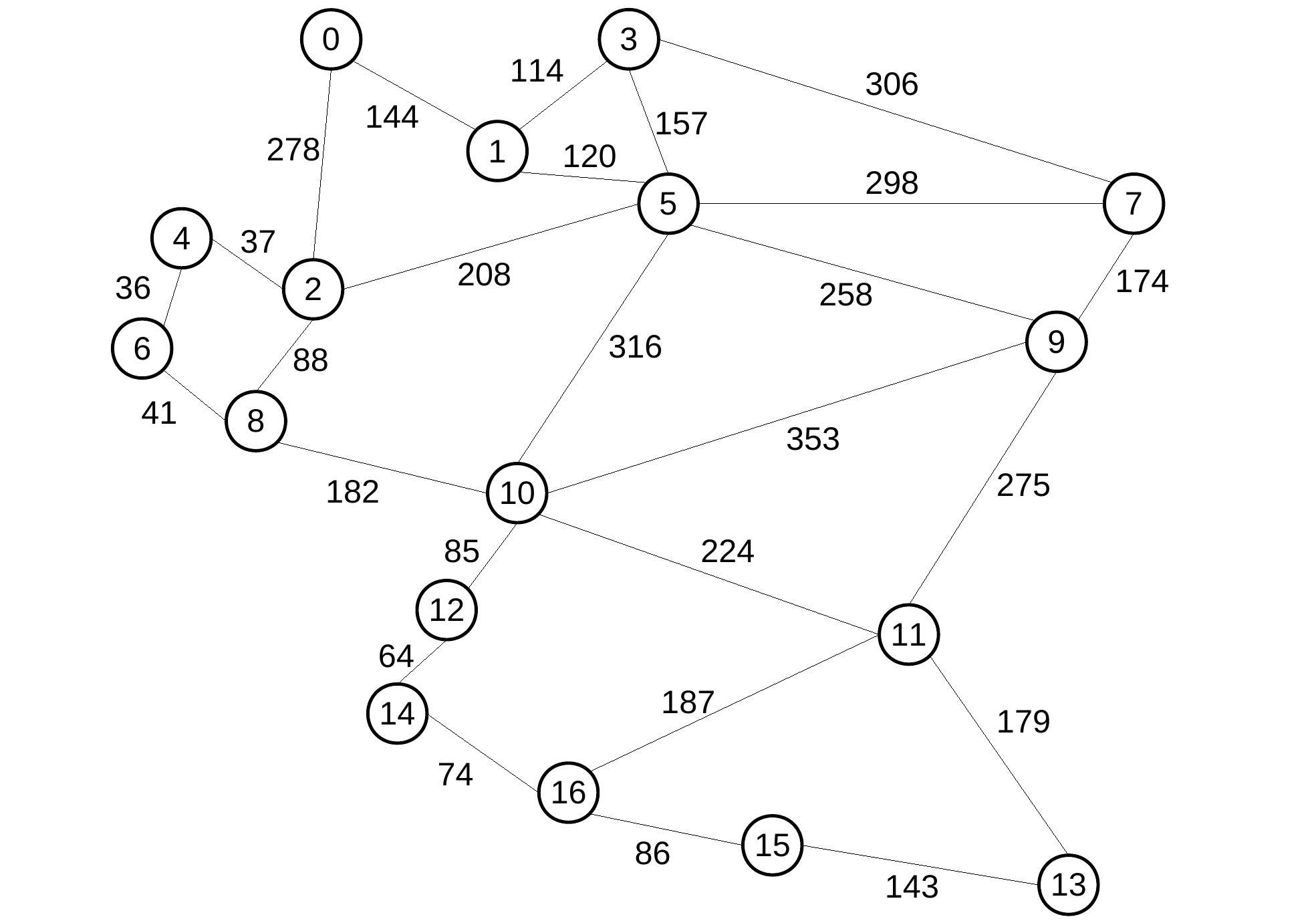}
		\caption{GERMAN topology.}\vspace{-6ex}
		\label{fig:german}
	\end{center}
\end{figure}

The network topologies considered are USA ($24$ nodes and $43$ bidirectional links) and GERMAN ($27$ nodes and $26$ bidirectional links), shown in Figures~\ref{fig:usanet} and \ref{fig:german}, respectively.
These topologies were used to exemplify two realistic yet distinct scenarios, one representing continental size dimensions (USA) and the other one with shorter dimensions (GERMAN).
Note that these scenarios will force the resource allocation solutions to adapt to a different range of modulation formats given their transparent reach.
%
\vspace{-2ex}

\subsection{Average Use of Modulation}\label{sec:mod}

In order to better illustrate the constraints of the different topology scenarios, we present the average modulation techniques usage (Table~\ref{tab:modulatio1}).
This metric indicates the percentage of lightpaths allocated per modulation level, it is an average value for all network loads. 
There was a small variation among loads, about $2\%$ of mean standard deviation for all schemes.
As can be observed in Table~\ref{tab:modulatio1}, more spectrally efficient modulations are rarely used in the USA topology, what was expected given that most of its links length are longer or closer to the modulation reach. On the other hand, in the GERMAN topology the less efficient modulations are practically not used.
The results also evinces that the DMMAS always take advantage of the most spectrally efficient modulation techniques provided by the network transponders, specially for short distances topologies, the same does not happens with the other schemes.
%
\vspace{-2ex}
\begin{table}[!h]
	\caption{Average use of modulation techniques ($\%$).}\vspace{-3ex}
	\label{tab:modulatio1}
	\begin{center}
		\small
		\begin{tabular}{|l|c|c|c|c|c|}
			\hline
			& \textbf{Modulation} & \textbf{mAdap} & \textbf{AMMS} & \textbf{EEMS} & \textbf{DMMAS} \\\hline
			\parbox[t]{2mm}{\multirow{6}{*}{\rotatebox[origin=c]{90}{\textbf{USA}}}}
			& \textit{BPSK}  & 24.00 & 0.07 & 23.65 & 0.16 \\\cline{2-6}
			& \textit{QPSK}  & 42.07 & 2.29 & 56.47 & 4.79 \\\cline{2-6}
			& \textit{8QAM}  & 21.66 & 13.37 & 10.82 & 21.27 \\\cline{2-6}
			& \textit{16QAM}  & 11.62 & 84.03 & 8.90 & 73.18 \\\cline{2-6}
			& \textit{32QAM}  & 0.33 & 0.11 & 0.05 & 0.45 \\\cline{2-6}
			& \textit{64QAM}  & 0.33 & 0.13 & 0.11 & 0.15 \\\hline
			\hline
			\parbox[t]{2mm}{\multirow{6}{*}{\rotatebox[origin=c]{90}{\textbf{GERMAN}}}}
			& \textit{BPSK}  & -- & -- & -- & -- \\\cline{2-6}
			& \textit{QPSK}  & -- & -- & 27.97 & -- \\\cline{2-6}
			& \textit{8QAM}  & -- & -- & -- & -- \\\cline{2-6}
			& \textit{16QAM}  & 40.26 & 40.26 & 59.93 & 14.42 \\\cline{2-6}
			& \textit{32QAM}  & 41.01 & 41.01 & 5.88 & 50.24 \\\cline{2-6}
			& \textit{64QAM}  & 18.72 & 18.72 & 6.22 & 35.34 \\\hline
		\end{tabular}\vspace{-8ex}
	\end{center}
\end{table}

\subsection{Bandwidth Blocking Ratio}\label{sec:bbr}

\begin{figure}[ht]
	\begin{center}
		\includegraphics[width=0.4\textwidth]{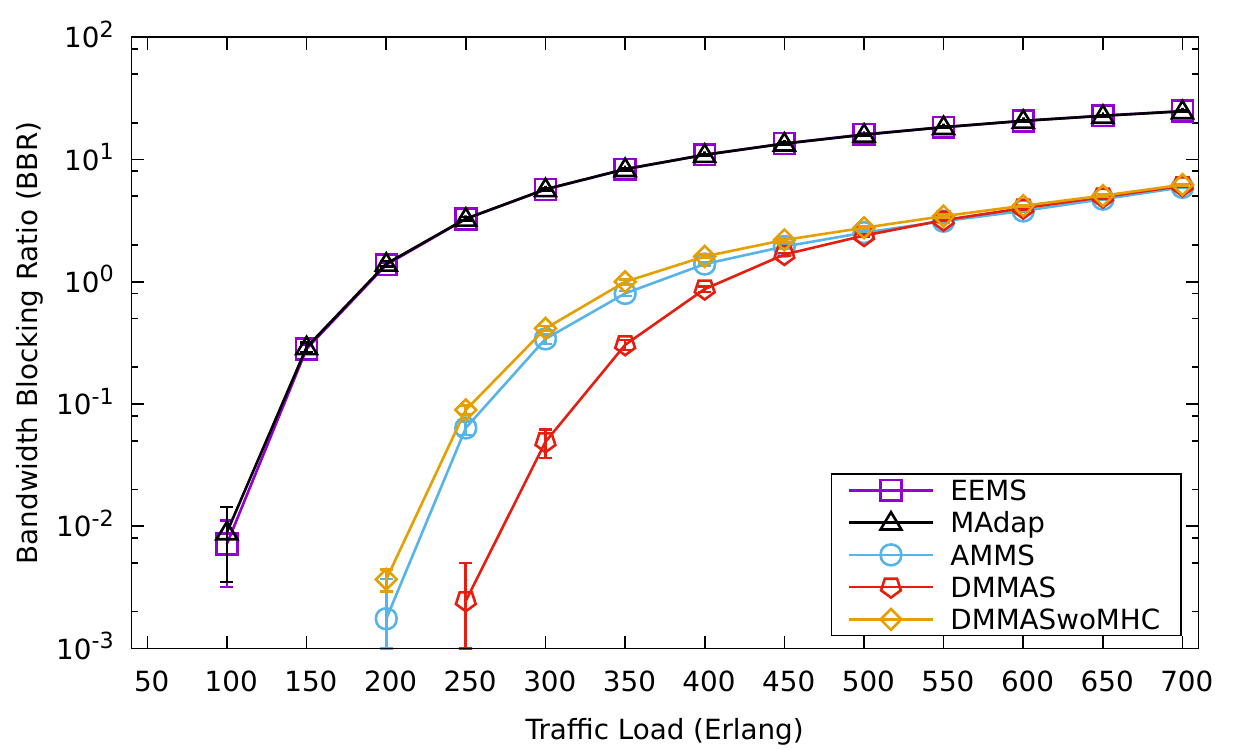}\vspace{-2ex}
		\caption{Bandwidth Blocking Ratio (BBR) for USA topology.}\vspace{-4ex}
		\label{fig:bbrUSANET}
	\end{center}
\end{figure}

\begin{figure}[ht]
	\begin{center}
		\includegraphics[width=0.4\textwidth]{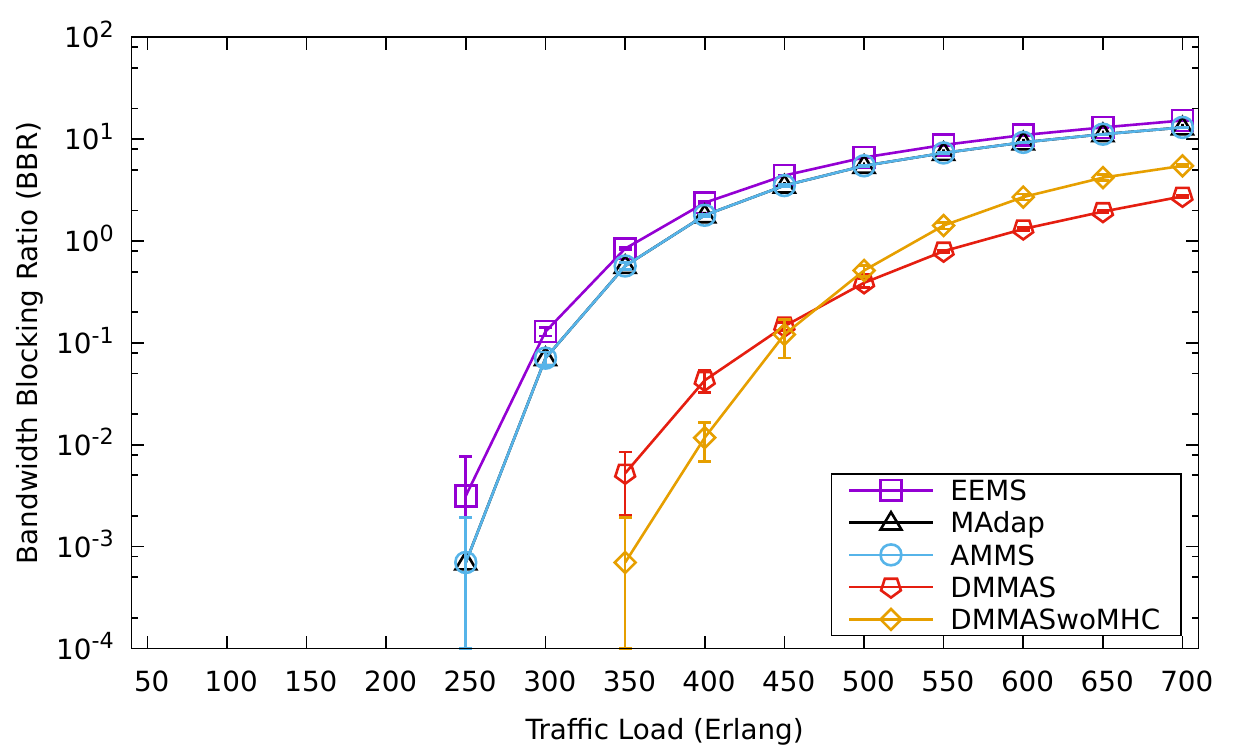}\vspace{-2ex}
		\caption{Bandwidth Blocking Ratio (BBR) for GERMAN topology.}\vspace{-6ex}
		\label{fig:bbrGERMAN}
	\end{center}
\end{figure}

The BBR (Figures~\ref{fig:bbrUSANET} and~\ref{fig:bbrGERMAN}) reflects the ratio of blocked bandwidth, higher values meaning more bandwidth is being blocked, so lower values are desirable.
We observe that the BBR performance in DMMAS schema achieve better performance than the second best scheme (AMMS), on average $77\%$ in USA and $86\%$ in GERMAN.
It can also be observed that for low and medium loads the DDMAS obtains gains of up to two orders of magnitude.
The gains in relation to the \textit{mAdap} and EEMS schemes are expected and already explored in~\cite{MBM} and~\cite{AMMS}. Since the use of multi-hop provides more opportunities to accept requests by relaxing the spectrum continuity and transmission distance constraints of the RMLSA problem.
In relation to AMMS, the DMMAS scheme brings two advantages: \textit{SpecEff-Module} and the dynamic $MHC$.
%
%
The benefits of dynamic MHC can be seen when compared to DMMASwoMHC. In the USA topology results, DMMASwoMHC becomes very similar to AMMS, due to the larger distances in that network. For the GERMAN topology the DMMASwoMHC presents good results for lower loads because it abuses on the number of OEO hops making its lightpaths very short allowing the utilization of the most efficient modulation formats, for higher loads this greedy behavior pays a price on blocking.
%
\vspace{-1ex}

\subsection{Average Number of Virtual Hops per request}\label{sec:vt}

\begin{figure}[ht]
	\begin{center}
		\includegraphics[width=0.4\textwidth]{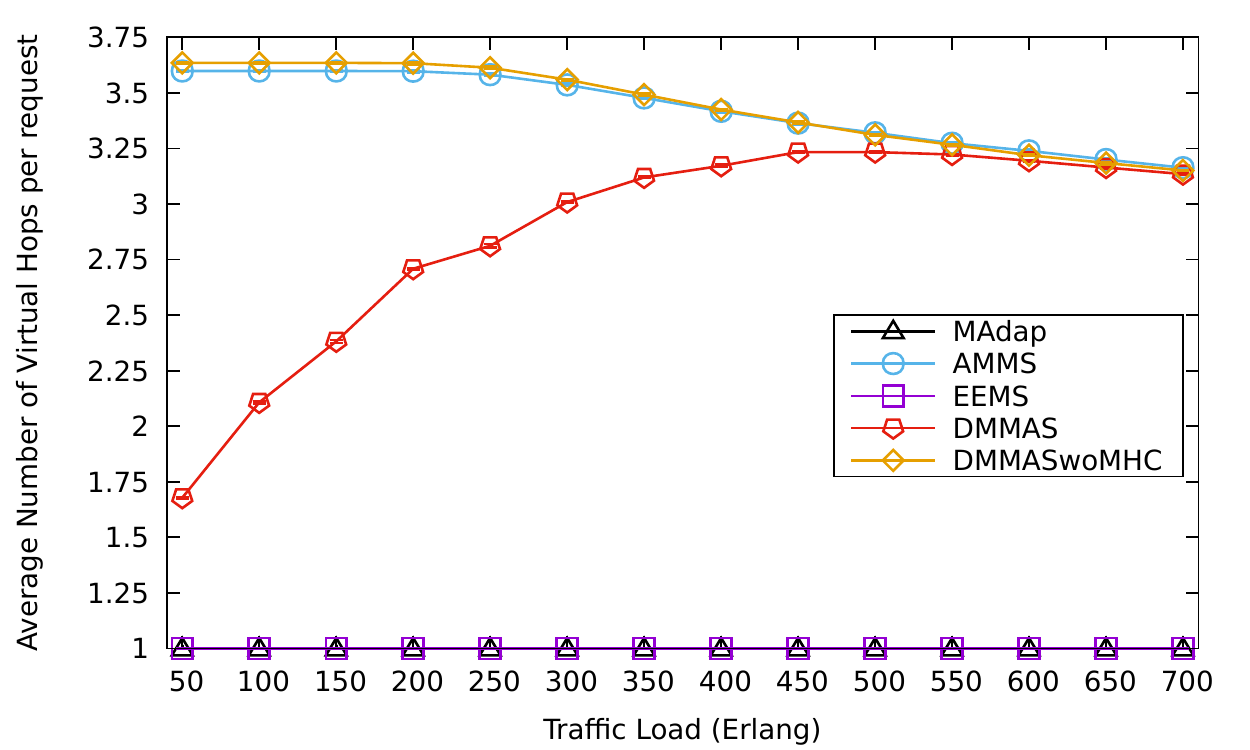}\vspace{-2ex}
		\caption{Average Number of Hops in the Virtual Topology per Request for USA topology.}\vspace{-4ex}
		\label{fig:vtUSANET}
	\end{center}
\end{figure}

\begin{figure}[ht]
	\begin{center}
		\includegraphics[width=0.4\textwidth]{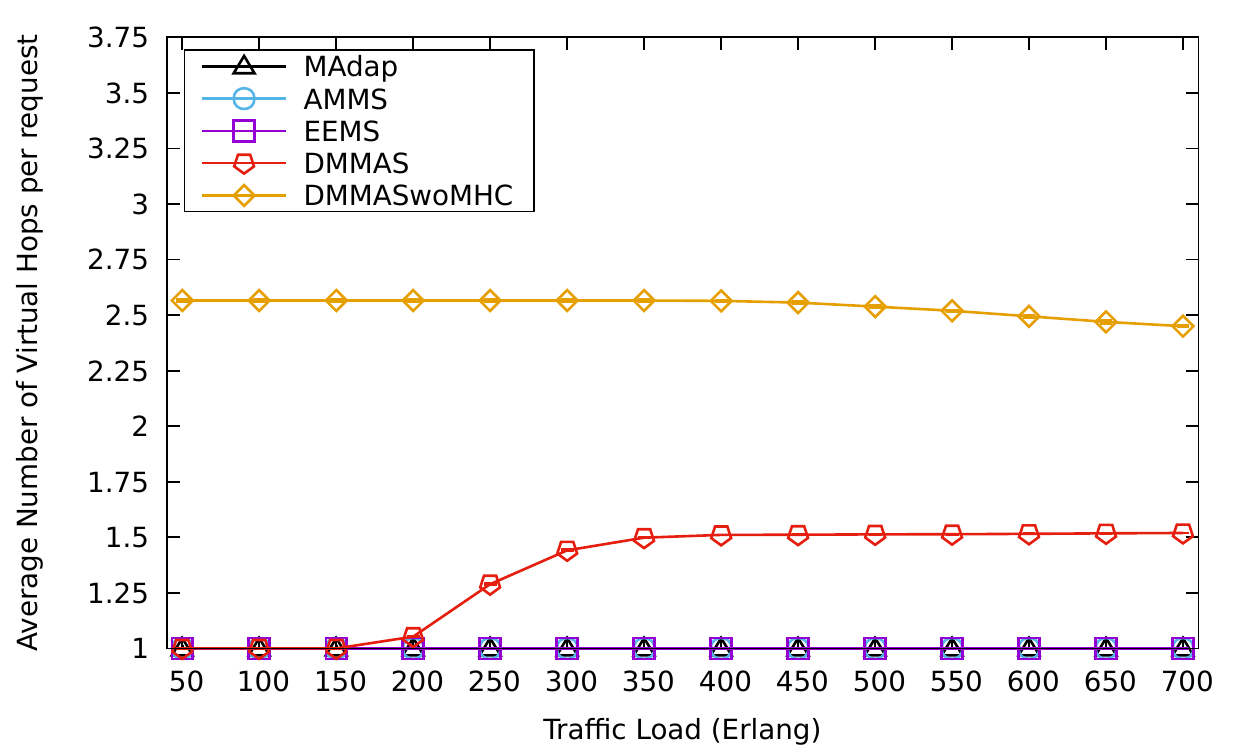}\vspace{-2ex}
		\caption{Average Number of Hops in the Virtual Topology per Request for GERMAN topology.}\vspace{-6ex}
		\label{fig:vtGERMAN}
	\end{center}
\end{figure}

The average number of virtual hops (Figures~\ref{fig:vtUSANET} and~\ref{fig:vtGERMAN}) indicates how many lightpaths (OEO conversions) were used to serve the traffic demands.
The algorithms under the \textit{mAdap} and EMMS schemes keeps in any traffic load an expected value of one hop, because they employ single-hop routing. 
For the AMMS scheme, a similar behavior is observed in the GERMAN topology, although it is a multi-hop scheme, this occur due to the short distances in this topology, leading the value of its static $MHC \rightarrow 1$~\cite{AMMS}.
For the USA topology, AMMS obtained the average virtual hop between $3$ and $3.75$ per request, close to the DMMASwoMHC.
%
%
%
%
This shows that in scenarios with large distances, such as the USA, the AMMS MHC is not effective, behaving like a solution without any limitation, such as DMMASwoMHC.
On the other hand, in shorter networks such as GERMAN, the AMMS behaves like a single-hop mechanism, such as mAdap and EEMS.
In contrast, in DMMAS the $MHC$ is dynamically calculated, based on the network fragmentation index.
This empowers the scheme to use more hops when resources become scanty, better adapting to the network state.
In GERMAN topology, the number of virtual hops proposed by DMMAS solutions is low, varying from $1$ to $1.5$ hops on average.
In USA topology, the DMMAS proposes solutions with more hops, reaching $3.25$ hops as the load increases.
These results show that the dynamic MHC proposed by DMMAS avoids meeting traffic requests with an excessive number of OEO conversions when the load is low and limits the unnecessary multi-hop routing, thus contributing significantly to the reduction of network latency.

\subsection{External Fragmentation ratio}\label{sec:frag}

\begin{figure}[ht]
	\begin{center}
		\includegraphics[width=0.4\textwidth]{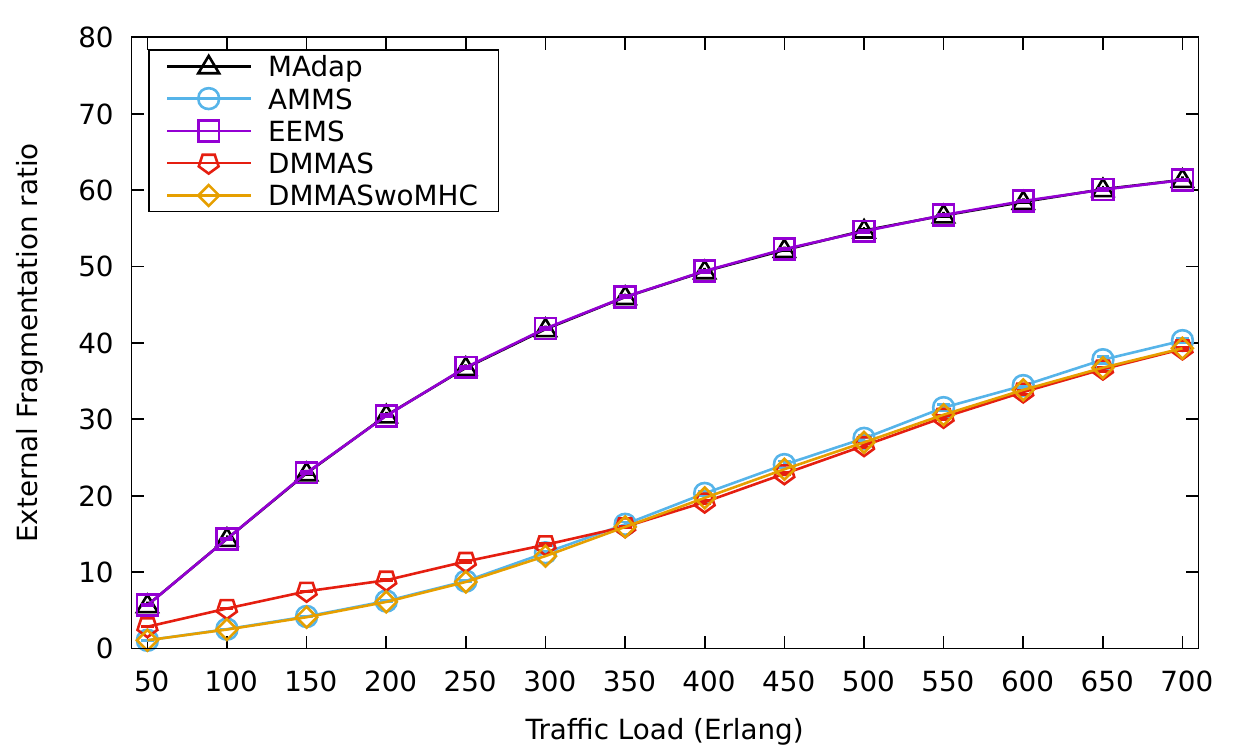}
		\caption{External Fragmentation ratio for USA topology.}\vspace{-4ex}
		\label{fig:fragUSANET}
	\end{center}
\end{figure}

\begin{figure}[ht]
	\begin{center}
		\includegraphics[width=0.4\textwidth]{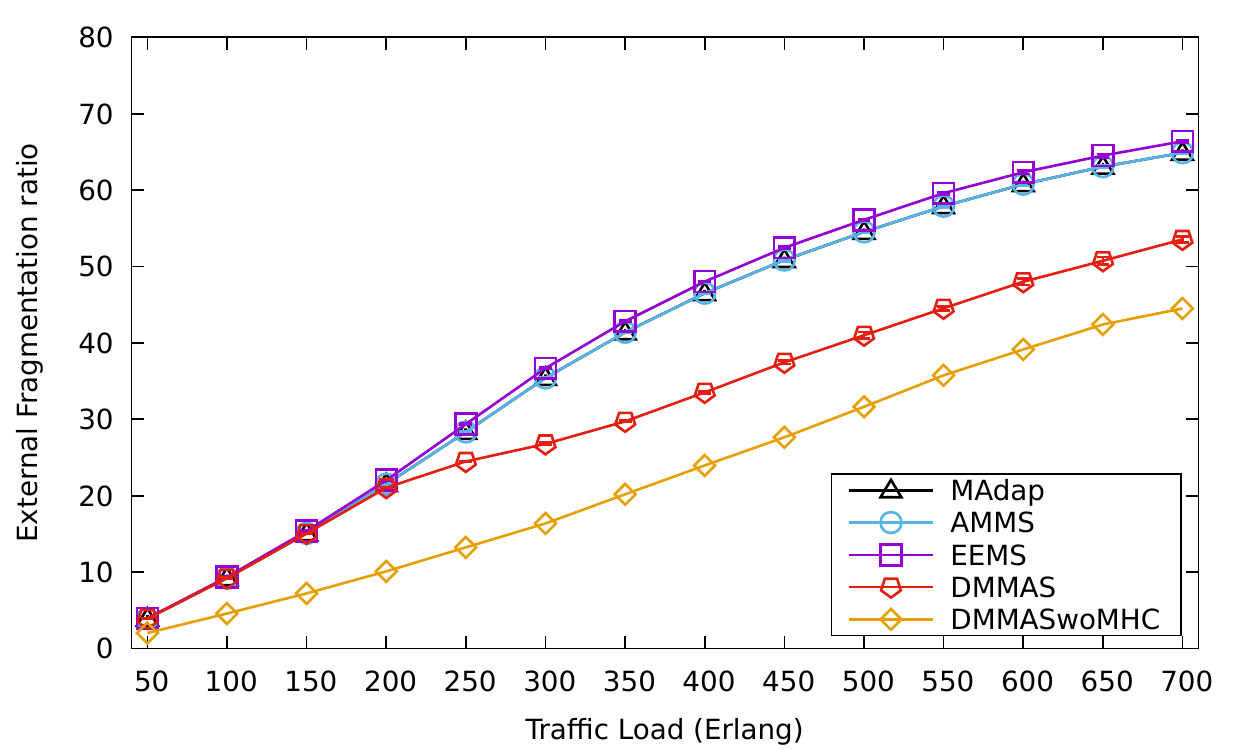}
		\caption{External Fragmentation ratio for GERMAN topology.}\vspace{-4ex}
		\label{fig:fragGERMAN}
	\end{center}
\end{figure}

The external fragmentation ratio (Figures~\ref{fig:fragUSANET} and~\ref{fig:fragGERMAN}) measures the average external fragmentation level of all network links during the simulation time, according to Eq.~\ref{eq:fextRosa}.
High levels represent a network with a fragmented spectrum and consequently an inefficient utilization of spectrum resources.
We can observe that for GERMAN topology the fragmentation rate of the DMMAS scheme is lower than the other schemes, with gains of up to $52\%$.
In the USA topology the fragmentation rate presented by our schemes was very close to the AMMS, an average of $19\%$. For the other schemes the DMMAS presents a reduction of up to $55\%$.
%
%
This is because DMMAS uses more efficient modulation levels, which carry more bits per symbol, and in turn form smaller channels with higher capacity.
Narrower channels impact less the fragmentation of the spectrum.

\subsection{Effective Energy Efficiency}\label{sec:energy}

\begin{figure}[ht]
	\begin{center}
		\includegraphics[width=0.4\textwidth]{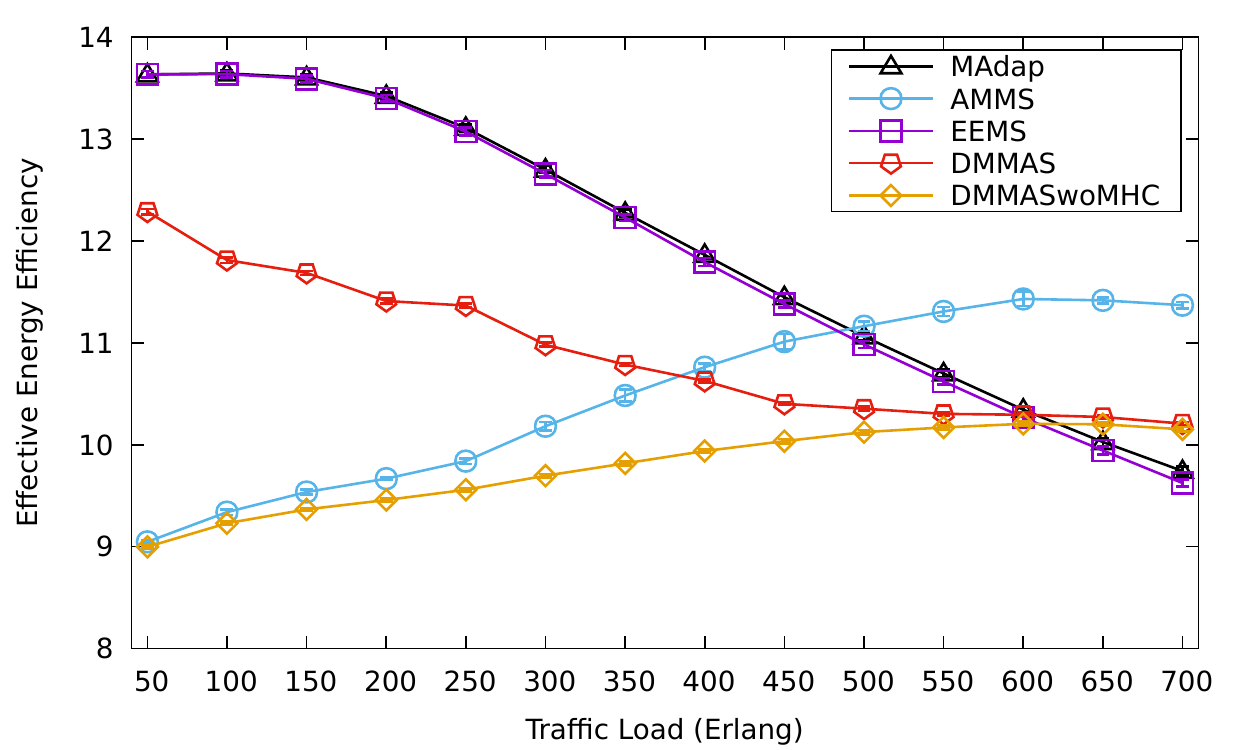}
		\caption{Effective Energy Efficiency for USA topology.}
		\label{fig:eeeUSANET}
	\end{center}
\end{figure}

\begin{figure}[ht]
	\begin{center}
		\includegraphics[width=0.4\textwidth]{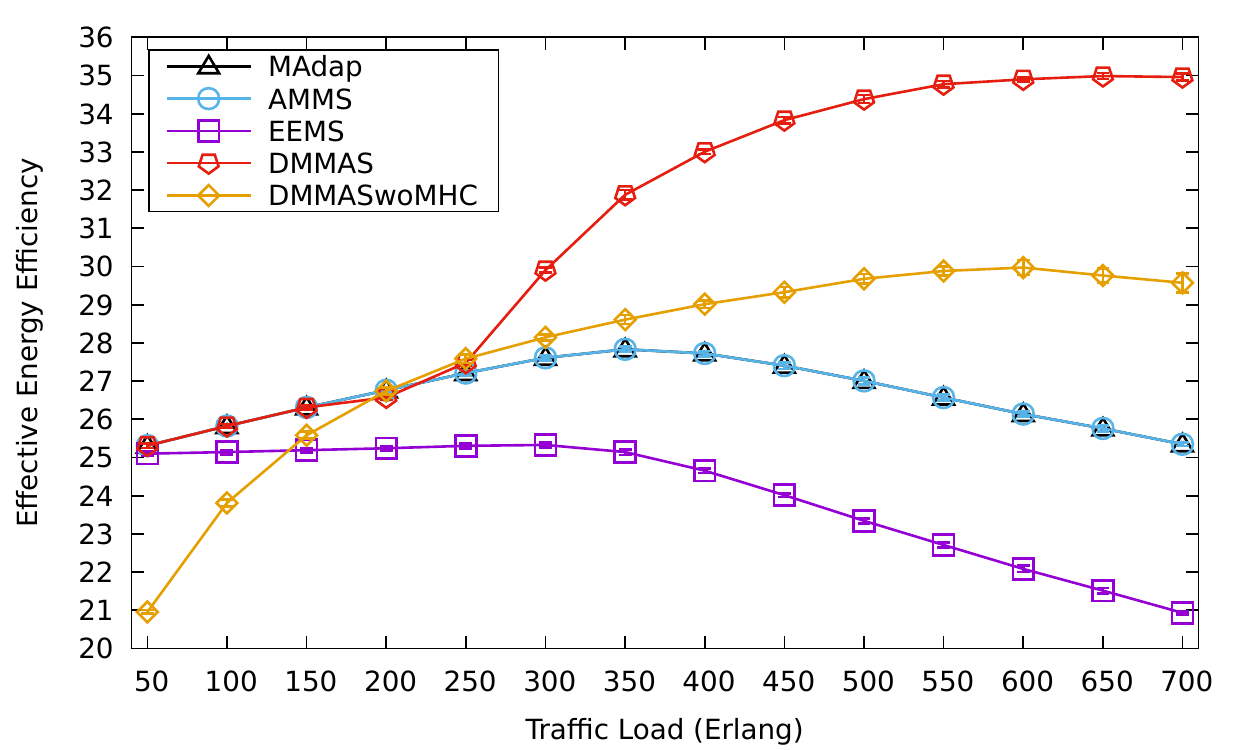}
		\caption{Effective Energy Efficiency for GERMAN topology.}
		\label{fig:eeeGERMAN}
	\end{center}
\end{figure}

In order to better evaluate the TE approach, we propose a new metric based on the product between the energy efficiency (Eq.~\ref{eqn:ee}) and the bandwidth blocking ratio\footnote{$BBR = \{x \in {\rm I\!R} : 0 \leq x \leq 1 \}$} of the algorithm, called effective energy efficiency, as shown in Eq.~\ref{eqn:eee}.
\begin{equation}\label{eqn:eee}
	\displaystyle EEE = EnEff(bits/Joule) \times (1 - BBR^{\textrm{\footnote[3]{}}})
\end{equation} 
%
%
This new metric (Figures~\ref{fig:eeeUSANET} and~\ref{fig:eeeGERMAN}) represents the trade-off between energy efficiency and network efficiency.
%
Higher values mean a more effective approach.
It is important to note that increasing the load does not increase the amount of data served. 
%
The relationship between single-hop and multi-hop routing in the context of energy efficiency is not yet well explored in the literature.
Using more segments in the RMLSA solution causes the use of more BVTs per request which causes a higher energy expenditure.
However, a multi-hop solution leads to the allocation of smaller segments which allow the use of more efficient modulation formats, thus using less spectrum slots.
Although the use of more efficient modulation levels naturally increase the energy consumption in transponders, this relation is not linear, as can be seen in Table~\ref{tb:1}. On the other hand, the impact of using less slots (channel size) directly affect the power consumption.
Therefore, the ideal trade-off is a balance between number of hops (use of BVTs) and the level of modulation used in the solution.

In USA topology, the DMMAS scheme provides effective energy efficiency of up to $5\%$ between the loads of 50 to 350 Erlangs in relation to the AMMS scheme.
For lighter loads, the \textit{mAdap} and EEMS schemes obtained better results, up to $8\%$.
This is due to the fact that these schemes do not use multi-hop solutions and therefore use fewer BVTs in the solution.
In contrast, in the GERMAN topology, the DMMAS presents gains of up to $40\%$ in relation to the schemes of the literature.
The short distances in that topology favors the performance of our scheme because the use of more efficient modulation levels generates a smaller power footprint.
%
%

\section{Conclusion}\label{sec:conclusion}

We have proposed a new approach to address the RMLSA problem based on new distance-adaptive modulation scheme which allows the use of any RSA algorithm on solving the RMLSA problem.
The DMMAS extends the adaptive modulation schemes of the literature by dynamically assigning the use of multi-hop routing through the network fragmentation index.
In addition, a mechanism is proposed that uses different modulation levels to meet the same traffic demand, using modulation levels appropriate to the size of the multi-hop segments.

To demonstrate the effectiveness of the adaptive modulation scheme proposed, we compared three adaptive modulation schemes from the literature, \textit{mAdap}, AMMS and EEMS, in two distinct network scenarios.
Simulation results demonstrated that the use of the adaptive modulation scheme proposed provides gains varying from 2 orders of magnitude to $86\%$ for higher loads in the bandwidth blocking rate, also providing gains in spectrum fragmentation by up to $55\%$ in the network.
Moreover, those gains were achieved without over exploring OEO conversions, given the controlled multi-hop technique employed.
We observe that the DMMAS scheme also increases the effective energy efficiency up to $40\%$, resulting in a better use of network resources.
As a future work, we intend to explore the physical impairments of EON, taking into account the SNR (signal-to-noise ratio) of the optical circuit rather than the transmission distance, in order to guarantee an acceptable QoT throughout the network.



\bibliographystyle{IEEEtran}
\bibliography{bibliografia}

%








\end{document}